\documentclass[3p,12pt]{elsarticle}
\usepackage[utf8]{inputenc}

\usepackage{amssymb}
\usepackage{amsmath}
\usepackage{nicefrac}
\usepackage{fancyvrb}

\usepackage{lineno}

\biboptions{sort&compress}

\newcounter{bla}

\journal{}

\usepackage{bm}
\usepackage{relsize}

\usepackage[labelformat=simple]{subcaption}
\usepackage{tabularx}
\usepackage{miller}
\usepackage{microtype}
\usepackage{xspace}
\usepackage{makecell} 

\usepackage{hyperref}
\hypersetup{
	colorlinks=true,
	linkcolor=black,
	citecolor=black,
	filecolor=black,
	urlcolor=black,
}

\usepackage{algorithm}
\usepackage{algpseudocode}
\usepackage{siunitx}
\usepackage[dvipsnames]{xcolor}
\newcommand{\etal}{\textit{et al.}\ }
\newcommand{\fire}{\textsc{fire}\xspace}
\usepackage{scalefnt} 
\newcommand\smallerf[2][0.85]{{\scalefont{#1}#2}} 
\newcommand{\fireo}{\textsc{fire}~\smallerf[0.76]{2.0}\xspace} 
\newcommand{\imd}{\textsc{imd}\xspace}
\newcommand{\lmp}{\textsc{lammps}\xspace}
\newcommand{\lmpv}{12 Dec 2018}

\newcommand{\gromacs}{\textsc{gromacs}\xspace}
\newcommand{\dlpoly}{\textsc{dl\_poly}\xspace}
\newcommand{\eon}{\textsc{eon}\xspace}
\newcommand{\ase}{\textsc{ase}\xspace}
\newcommand{\lbfgs}{\textsc{l-bfgs}\xspace}
\newcommand{\neb}{\textsc{neb}\xspace}
\newcommand{\cgrad}{\textsc{cg}\xspace}
\newcommand{\sdesc}{\textsc{sd}\xspace}
\newcommand{\artmethod}{\textsc{art}\xspace}
\newcommand{\md}{\textsc{md}\xspace}
\newcommand{\eam}{\textsc{eam}\xspace}
\newcommand{\meam}{\textsc{meam}\xspace}
\newcommand{\sw}{\textsc{sw}\xspace}
\newcommand{\bks}{\textsc{bks}\xspace}
\newcommand{\ff}{\textsc{ff}\xspace}
\newcommand{\fcc}{\textsc{fcc}\xspace}
\newcommand{\hcp}{\textsc{hcp}\xspace}

\newcommand{\pbc}{\textsc{pbc}\xspace}
\newcommand{\fnorm}{\texttt{f2norm}\xspace}
\newcommand{\nlastpneg}{\ensuremath{N_{P \leq 0}}}
\newcommand{\nlastpnegmax}{\ensuremath{N_{P \leq 0, max}}}
\newcommand{\nlastppos}{\ensuremath{N_{P    > 0}}}

\begin{document}

\begin{frontmatter}

\title{Assessment and optimization of the fast inertial relaxation engine (\fire) for energy minimization in atomistic simulations and its implementation in \lmp}

\author[a,b,f]{Julien Gu\'enol\'e \corref{author}}
\author[c,d]{Wolfram G. N\"ohring}
\author[b]{Aviral Vaid}
\author[b]{Fr\'ed\'eric Houll\'e}
\author[b]{Zhuocheng Xie}
\author[e,b]{Aruna Prakash}
\author[b]{Erik Bitzek}

\cortext[author] {Corresponding author.\\\textit{E-mail address:} julien.guenole@univ-lorraine.fr}
\address[a]{Institute of Physical Metallurgy and Materials Physics, RWTH Aachen University, Germany}
\address[b]{Department of Materials Science and Engineering, Institute I, Friedrich-Alexander-Universit\"at Erlangen-N\"urnberg (FAU), Martensstr. 5, 91058 Erlangen, Germany}
\address[f]{LEM3, CNRS – Universit\'e de Lorraine – Arts et M\'etiers ParisTech, 7 rue F\'elix Savart, 57070 Metz, France}
\address[c]{Department of Microsystems Engineering, University of Freiburg, 79110 Freiburg, Germany}
\address[d]{Institute of Mechanical Engineering, \'Ecole Polytechnique F\'ed\'erale de Lausanne, EPFL STI IGM  Station 9, CH-1015 Lausanne}
\address[e]{Micromechanical Materials Modelling (MiMM), Institute of Mechanics and Fluid Dynamics, TU Bergakademie Freiberg, D-09599 Freiberg, Germany}

\begin{abstract}
In atomistic simulations, pseudo-dynamics relaxation schemes often exhibit better performance and accuracy in finding local minima than line-search-based descent algorithms like steepest descent or conjugate gradient.
Here, an improved version of the fast inertial relaxation engine (\fire) and its implementation within the open-source atomistic simulation code \lmp is presented. It is shown that the correct choice of time integration
scheme and minimization parameters is crucial for the performance of \fire.
\end{abstract}


\end{frontmatter}


\section{Introduction}
Numerical optimization \cite{Fletcher2013,Nocedal2006} is of utmost importance in almost every field of science and engineering.
It is routinely used in atomistic simulations in condensed matter physics, physical chemistry, biochemistry, and materials science. 
There, the optimized quantity is usually the potential energy $E(\mathbf{x})$, for a given interatomic interaction model \cite{Schlick2010}.
\emph{Minimizing} $E(\mathbf{x})$
with respect to the atomic coordinates $\mathbf{x}$  yields 0~K equilibrium structures and energies,
e.g., of defects.
Minimum energy configurations can, furthermore, be used as initial states for subsequent molecular dynamics (\md) simulations or normal-mode analyses~\cite{Umeno10PRB}.
Energy minimization is also used to determine the stability of structures under load. 
Two typical examples are the computation of the Peierls stress required for dislocation glide \cite{Bitzek2005}, and the determination of the critical stress intensity factor required for crack propagation \cite{Moller2014b}. 
Other uses of energy minimization methods in atomistic simulations include the search for transition states, e.g.\ by the nudged-elastic-band (\neb) method \cite{Sheppard08JCP}, or the detection of transitions in accelerated \md methods like parallel-replica dynamics or hyperdynamics \cite{perez2009accelerated}.

Most atomistic simulation packages like \lmp{} \cite{Plimpton95JCP}, \gromacs \cite{Abraham2015GROMACS}, \imd \cite{Stadler1997IMD}, \dlpoly \cite{Smith2002DL_POLY}, \eon \cite{Chill2014} or \ase \cite{Larsen2017} implement line-search-based descent algorithms like Steepest Descent (\sdesc) or Conjugated Gradient (\cgrad), as well as damped-dynamics methods like Microconvergence \cite{Beeler1983MIK}, Quickmin \cite{Sheppard2008quickmin} and the Fast Inertial Relaxation Engine (\fire) \cite{Bitzek06PRL}.
More complex algorithms including Quasi-Newton methods like the highly-efficient Limited-memory Broyden-Fletcher-Goldfarb-Shanno (\lbfgs) approach that involve the computation of the Hessian \cite{Nocedal2006} are mostly used in ab-initio simulations and are not as widely implemented in atomistic simulations packages as the aforementioned Hessian-free algorithms.

\fire is often used in atomistic simulations of mechanical properties of metals and alloys 
\cite{Moller2014b,Nogaret2010}, ceramics \cite{Pastewka13PRB}, polymers \cite{Riggleman2010}, carbon allotropes \cite{Amsler2012}, amorphous materials \cite{Singh2013} and granular media \cite{Dagois-Bohy2012}, as well as in simulations related to catalysis \cite{Brogaard2014} or docking \cite{Fanfrlik2010}.
The strict adherence to force minimization in \fire makes it ideally suitable for critical point analysis in translational invariant systems like for the determination of the Peierls stress of a dislocation \cite{Bitzek2005,Bitzek06PRL}, where line-search-based descent algorithms often fail. 
Furthermore, \fire has been shown to be a convenient algorithm for mapping basins of attraction, as it avoids unusual pathologies like disconnected basins of attraction that can appear, e.g., using the \lbfgs method \cite{Asenjo2013}.
\fire was also shown to be a fast and computationally efficient minimizer for \neb \cite{Sheppard08JCP}, as well as for the activation-relaxation technique (\artmethod) \cite{Machado-Charry2011}.

Here, we study the influence of the numerical integration scheme and the choice of parameters set (mixing coefficient, initial timestep, maximum timestep, etc.) on the efficiency of \fire\ for different scenarios.
We furthermore suggest a modification of the \fire\ algorithm to improve its efficiency and describe our implementation of this modified version \fireo{} in the atomistic simulation code \lmp~\cite{Plimpton95JCP}. 

\section{The algorithms}

\subsection{\fire}
\label{sec:algo_fire}

Consider a system of $N$ particles with coordinates $\mathbf{x}\equiv(x_1,x_2,\dots,x_{3N})$ and mass $m$. The potential energy $E(\mathbf{x})$ depends only on the relative positions of the particles and can thus be envisioned as a 
$(3N-6)$-dimensional surface or ``landscape''. The principle of \fire is to perform dynamics which allow only for downhill motion on this landscape, with the acceleration
\begin{equation}
\label{equ_velocity}
\mathbf{\dot v}(t) = \frac{\mathbf{F}(\mathbf{x}(t))}{m} - \gamma(t) |\mathbf{v}(t)| 
\left(\mathbf{\hat v}(t) - \mathbf{\hat F}(\mathbf{x}(t)) \right).
\end{equation}
Here, $t$ denotes time,  $\mathbf{v}(t)$ the velocity of the particles ($\mathbf{v}(t)\equiv \mathbf{\dot x}(t)$), $\mathbf{F}(\mathbf{x}(t))$ the force acting on them, i.e., the gradient of the potential energy ($\mathbf{F}(\mathbf{x}(t)) = -\nabla E(\mathbf{x}(t))$), and 
$\gamma(t)$ a scalar function of time. Boldface quantities denote vectors, hats indicate unit vectors, and $\vert\dots\vert$ is the Euclidean norm of the enclosed vector. The first term on the right hand side in equation~\ref{equ_velocity}
represents regular Newtonian dynamics. The effect of the second term  is to reduce the angle between
$\mathbf{v}(t)$ and $\mathbf{F}(\mathbf{x}(t))$, which is the 
direction of steepest descent at $\mathbf{x}(t)$.  Uphill motion is avoided
by computing  the power  $P(t)= \mathbf{F}(\mathbf{x}(t)) \cdot \mathbf{v}(t)$ and setting the velocity to zero whenever $P(t)\leq0$.
It was shown that combining equation~\ref{equ_velocity} with an adaptive time stepping scheme yields a simple and competitive optimization algorithm \cite{Bitzek06PRL}.
In practice, equation~\ref{equ_velocity} is implemented by ``mixing'' $\mathbf{v}(t)$ and $\mathbf{F}(\mathbf{x}(t))$, 
using an adaptive mixing factor $\alpha(t)\in[0, 1]$.
The algorithm can then be written as proposed in Algorithm~\ref{algo:firealgo}.

\begin{algorithm}
\caption{\fire}\label{algo:firealgo}
\begin{algorithmic}[1]
\State Initialize $\mathbf{x}(t)$ and  $\mathbf{F}(\mathbf{x}(t))$
\State $\mathbf{v}(t) \gets 0$
\State $\alpha \gets \alpha_{\mathrm{start}}$
\State $\Delta t  \gets \Delta t_\mathrm{start}$ 
\State $\nlastppos\gets 0$
\For{$i\gets 1, N_\mathrm{max}$}
    \State $P(t) \gets \mathbf{F}(\mathbf{x}(t)) \cdot \mathbf{v}(t)$
    \If{$P(t) > 0$}
        \State $\nlastppos \gets \nlastppos+1$
        \State $\mathbf{v}(t) \gets (1 - \alpha)  \mathbf{v}(t) + \alpha  \mathbf{F}(\mathbf{x}(t)) |\mathbf{v}(t)|/\vert\mathbf{F}(\mathbf{x}(t))\vert$
        \If{$\nlastppos>N_{delay}$}
            \State $\Delta t \gets \mathsf{min}(\Delta t f_{inc}, \Delta t_{max})$ 
            \State  $\alpha \gets\alpha f_\alpha$
        \EndIf
    \ElsIf{$P(t) \leq 0$}
        \State $\nlastppos\gets 0$
        \State $\mathbf{v}(t) \gets 0$
        \State $\Delta t \gets \Delta t f_{dec}$ 
        \State $\alpha \gets \alpha_{start}$
    \EndIf
    \State Calculate $\mathbf{x}(t+\Delta t)$, 
                     $\mathbf{v}(t+\Delta t)$,
                     $\mathbf{F}(\mathbf{x}(t+\Delta t))$,  
                     $E(\mathbf{x}(t+\Delta t))$
                     \Comment{MD integration}
    \State $t \gets t+\Delta t$
    \If{converged}
        \State \textbf{break}
    \EndIf
\EndFor
\State $\Delta t  \gets \Delta t_\mathrm{start}$ 
\end{algorithmic}
\end{algorithm}

\subsection{\fireo}

In Ref.~\cite{Bitzek06PRL}, it was suggested that \fire\ can be used in conjunction with any common \md integrator. However, \fire\ implements a variable time-stepping scheme to speed up the 
descent. Therefore, the integrator must be robust against a change of timestep during integration. For example, a simple \textit{Euler explicit} integration scheme is not suitable. Symplectic schemes like Euler semi-implicit (also called \textit{symplectic Euler}), Leapfrog or Velocity Verlet are more robust against varying  timesteps~\cite{Cromer81AJP,Verlet67PR,Donnelly05AJP}.
Similarly, the recent work by Shuang \etal highlighted the importance of a suitable integration scheme for \fire\cite{Shuang19CMS}.
The choice of an adequate integrator for \fireo\ will be presented and discussed in this manuscript.

An important principle of \fire~\cite{Bitzek06PRL} is to set the velocity to zero \textit{as soon as} $P(t)$ is not positive anymore, that is $P(t)<=0$. However, that is numerically impossible, leading to overshooting. Due to discrete time integration, the system will have already gone uphill before $P(t)<0$ is detected.  
One could correct overshoot by moving backwards for one entire step $\Delta t$ and then re-starting the motion at time $t - \Delta t$. This will undo the uphill motion as expected, but could keep the trajectory too far from where $P(t) = 0$. A less aggressive correction is to move backward for half a timestep ($0.5 \Delta t$). 

The algorithm of \fireo can be written as proposed in Algorithm~\ref{algo:fireoalgo}, with the modification from Algorithm~\ref{algo:firealgo} highlighted in blue.\footnote{Note that the time \textit{t} of \textbf{v}(\textit{t}) and \textbf{x}(\textit{t}) in algorithm~\ref{algo:fireoalgo} correspond to an MD integration with the Euler and Velocity Verlet methods. It has to be slightly adapted for the Leapfrog integration method, since the evaluation of \textbf{v} and \textbf{x} are not synchronized.}.

\begin{algorithm}
\caption{\fireo}\label{algo:fireoalgo}
\begin{algorithmic}[1]
\State Initialize $\mathbf{x}(t)$ and  $\mathbf{F}(\mathbf{x}(t))$
\State $\mathbf{v}(t) \gets 0$
\State $\alpha \gets \alpha_{\mathrm{start}}$
\State $\Delta t  \gets \Delta t_\mathrm{start}$ 
\State $\nlastppos\gets 0$
\State \textcolor{Blue}{$N_{P\leq0} \gets 0$}
\For{$i\gets 1, N_\mathrm{max}$}
    \State $P(t) \gets \mathbf{F}(\mathbf{x}(t)) \cdot \mathbf{v}(t)$
    \If{$P(t) > 0$}
        \State $\nlastppos \gets \nlastppos+1$
        \State $\nlastpneg \gets 0$
        \If{$\nlastppos>N_{delay}$}
            \State $\Delta t \gets \mathsf{min}(\Delta t f_{inc}, \Delta t_{max})$ 
            \State  $\alpha \gets\alpha f_\alpha$
        \EndIf
    \ElsIf{$P(t) \leq 0$}
        \State $\nlastppos \gets 0$
        \textcolor{Blue}{
        \State $\nlastpneg \gets \nlastpneg + 1$
        \If{$\nlastpneg > \nlastpnegmax$}
            \State \textbf{break}
        \EndIf
        \If{not (\texttt{initialdelay} and $i<N_\mathrm{delay}$)} 
            \If{$\Delta t f_{dec} \geq  \Delta t_{min}$}
                \State $\Delta t  \gets \Delta t f_{dec}$
            \EndIf
            \State $\alpha \gets  \alpha_{\mathrm{start}}$
        \EndIf
        \State $\mathbf{x}(t)\gets \mathbf{x}(t) - 0.5\Delta t \mathbf{v}(t)$ \Comment{Correct uphill motion}}
        \State $\mathbf{v}(t) \gets 0$
    \EndIf
    \State Calculate $\mathbf{x}(t+\Delta t)$, 
                     $\mathbf{v}(t+\Delta t)$,
                     $\mathbf{F}(\mathbf{x}(t+\Delta t))$,  
                     $E(\mathbf{x}(t+\Delta t))$
                     \Comment{MD integration and mixing (See algorithms \ref{algo:eulerexpl}-\ref{algo:verlet})}
    \State $t \gets t+\Delta t$
    \If{converged}
        \State \textbf{break}   
    \EndIf
\EndFor
\State $\Delta t  \gets \Delta t_\mathrm{start}$ 
\end{algorithmic}
\end{algorithm}

\section{Implementation in LAMMPS}

\label{sec:implementation_fire}

\subsection{Time integration scheme}

Historically, \fire\ has been developed for the \md code \imd~\cite{Stadler97IJMPC}, which implements a Leapfrog integrator for both dynamics and quenched-dynamics simulations. Thus, the published algorithm implicitly used Leapfrog, and the effect of this choice on \fire\ has not been addressed yet.
In the \md code \lmp~\cite{Plimpton95JCP}, \fire\ doesn't use the same \md integrator that is used for regular dynamics (Velocity Verlet), but a dedicated integrator. In the current implementation (\lmpv) this is the Explicit Euler method. Explicit Euler integration is not commonly used in classical \md, where  the requirement for energy conservation over long time periods suggests symplectic integrators~\cite{Donnelly05AJP,Tadmor11Book}. To investigate the influence of the integrator, we implemented  Euler Explicit (Algorithm~\ref{algo:eulerexpl}), Euler Semi-implicit (Algorithm~\ref{algo:eulerimpl}) and Velocity Verlet (Algorithm~\ref{algo:verlet}) methods. See \ref{sec:source} for the source code.

In addition, we also considered the Leapfrog (Algorithm~\ref{algo:leapfrog}) integration scheme which differs from Euler semi-implicit only in the initialization of velocities. Since the velocities are reset to zero at the beginning of the pseudo-dynamic run and also periodically during the run in \fireo, it turns out that both integrators are almost identical, as also confirmed by preliminary simulations. Therefore, the Leapfrog integrator is not considered for assessing \fireo in this manuscript\footnote{The source code of the Leapfrog integrator is present in the implementation of \fireo in \lmp for testing purposes only (See \ref{sec:source}). It is accessible by using the keyword \texttt{leapfrog} for the argument \texttt{integrator}, see Tab.~\ref{tab:parameters}}.

\subsection{Correcting uphill motion}

This correction is indicated in Algorithm~\ref{algo:fireoalgo}, and referred to as \texttt{halfstepback} in the \lmp\ implementation.

\subsection{Adjustments for improved stability}

The first adjustment consists of delaying the increase of $\Delta t$ and decrease $\alpha(t)$ for a few steps after $P(t)$ becomes negative. 
The second adjustment is to perform the mixing of velocity and force vectors ($\mathbf{v} \rightarrow (1 - \alpha)  \mathbf{v} + \alpha  \mathbf{\hat{F}}(\mathbf{x}) |\mathbf{v}|$) just before the last part of the time integration scheme, instead of at the beginning of the step. Note that this modification has no effect if \fire is used together with the Euler explicit integrator. 

\subsection{Additional stopping criteria}

An additional stopping criteria has been implemented in \fireo\ in order to avoid unnecessary looping, when it appears that further relaxation is impossible (stopping return value \texttt{MAXVDOTF} in \lmp). This could happen when the system is stuck in a narrow valley, bouncing back and forth from the walls but never reaching the bottom. The criterion is the number of consecutive iterations with $P(t) < 0$. Minimization is stopped if this number exceeds a threshold (\texttt{vdfmax} in the \lmp\ implementation).

We would like to comment on the force-based stopping criterion. While threshold defined for the minimization is usually not mentioned in the literature, the exact definition of the threshold is strongly related to the code.
\lmp uses the \fnorm criterion that corresponds to the Euclidean norm of the $3 \times N$ force vector.
Other codes might use less strict criteria, like the maximum force component acting on any atom, or the maximum force component per degree of freedom of the system. 
On overall and to compare the different degrees of relaxation that can be achieved, it is important to note that the \fnorm criterion considered by \lmp can be several order of magnitude stricter than the others. This has to be considered when comparing systems relaxed with different codes and the exact criterion should be reported in publications.

\section{Usage of \fireo in \lmp}
\label{Sec:usage}

Energy minimization in \lmp\ is performed with the command \texttt{minimize}. The type of minimization is set by \texttt{min\_style}, the default choice being the conjugate gradient method. \texttt{min\_style fire2} currently\footnote{\label{foot:fire}Please refer to the documentation of \lmp for the exact keyword enabling \fireo, or see \ref{sec:source}.} selects \fireo{}. The command \texttt{min\_modify} allows the user to tune parameters of the minimizations. The arguments, possible values, default value and description are listed in Tab.~\ref{tab:parameters}.
Below is an example of \fireo usage in \lmp (See \ref{sec:source} for accessing the source code):

\begin{table}[t]
	\centering
	\small
	\begin{tabular}{lll}
		\hline 
		Argument & Choice (default) & Description \\
		\hline 
		\texttt{integrator} & \makecell[l]{\texttt{eulerimplicit} \\ \texttt{eulerexplicit}\\ \texttt{verlet}\\
		(\texttt{eulerimplicit})} & Integration scheme \\
		\texttt{tmax} & \textit{float} 
		($10.0$) & The maximum timestep is  $\texttt{tmax}\times\Delta t_\mathrm{start}$\\
		\texttt{tmin} & \textit{float} 
		($0.02$) &  The minimum timestep is $\texttt{tmin}\times\Delta t_\mathrm{start}$ \\
		\texttt{delaystep} & \textit{integer} 
		($20$) & Number of steps to wait after $P<0$ before increasing $\Delta t$\\
		\texttt{dtgrow} & \textit{float} 
		($1.1$) & Factor by which $\Delta t$ is increased\\
		\texttt{dtshrink} & \textit{float} 
		($0.5$) & Factor by which $\Delta t$ is decreased\\
		\texttt{alpha0} & \textit{float} 
		($0.25$) & Coefficient for mixing velocity and force vectors\\
		\texttt{alphashrink} & \textit{float} 
		($0.99$) & Factor by which $\alpha$ is decreased\\
		\texttt{vdfmax} & \textit{integer}  
		($2000$) & Exit after \texttt{vdfmax} consecutive iterations with $P(t)<0$\\
		\texttt{halfstepback} & \texttt{yes}, \texttt{no}  
		(\texttt{yes}) & \texttt{yes} activates the inertia correction\\
		\texttt{initialdelay} & \texttt{yes}, \texttt{no}  
		(\texttt{yes}) & \texttt{yes} activates the initial delay in modifying $\Delta t$ and $\alpha$\\
		\hline 
	\end{tabular}
	\caption{Arguments of the command \texttt{min\_modify} in \lmp\ that define the parameters of the \fireo\ minimization method. Default values are in brackets.}
	\label{tab:parameters}
\end{table} 

\begin{Verbatim}[samepage=true]
#units metal
timestep 0.002
min_style fire2
min_modify integrator verlet tmax 6.0
minimize 0.0 1.0e-6 10000 10000
\end{Verbatim}

These commands instruct \lmp{} to 
perform energy minimization until \texttt{f2norm} falls below $10^{-6}$~eV/\AA\  or 10,000 force evaluations have been reached. Velocity Verlet integration is used and the maximum timestep is 
\SI{0.012}{\pico\second}.

\section{Assessing \fireo for typical applications in material science}

\subsection{Typical optimization problems in material science}

To assess the implementation of \fireo in \lmp, we use eight test cases (See section~\ref{sec:simulation-setups}) that address the following common problems in material science:
\begin{itemize}[--]
	\item Simultaneous relaxation of long range strain fields and short range disturbances (cases \ref{c:disl}, \ref{c:NPGpbc} and \ref{c:NPGnopbc}).	\item Relaxation of electrostatic interactions with short range rearrangements and atoms of different mass (case \ref{c:glass}).
	\item Relaxation of short and long range stress fields with a strongly directional atomic bonds (case \ref{c:SiVac}).
	\item Relaxation of a long range stress field of relatively low magnitude (case \ref{c:Mg}).
	\item Relaxation of systems with 3-body interactions (cases \ref{c:SiVac}, \ref{c:Mg} and \ref{c:Laves}).
	\item \neb calculations, i.e.\ simulatenous energy minimization of an ensemble of systems with 
	modified forces (cases \ref{c:AlVac} and \ref{c:Laves}). In case \ref{c:AlVac}, the 
	converged solution is closer to the initial guess than in case \ref{c:Laves}.
	
\end{itemize}

\subsection{The force fields}

The aforementioned tests rely on four different classes of force fields (\ff), which are described in the following and summarized in Tab.~\ref{tab:tests}.

The Embedded Atom Method (\eam{}) potential \cite{Daw84PRB, Foiles86PRB} is a widely used \ff in atomistic simulations of materials in general and of metals in particular~\cite{Rodney00PRB, Bachurin10AM, Prakash15AM, Prakash17Materials, Prakash17IJP, Zepeda-Ruiz17N, Chang19NC}. It is thus the primary \ff of the test cases. 
The \eam is a function of a two-body term and an ``embedding energy'', which is a functional of the local electron density. The latter is calculated based on contributions from radially symmetric electron density functions of atoms in the environment. Here, \eam is used for simulating Au and Al.

The Modified Embedded Atom Method (\meam) potential \cite{Baskes87PRL, Baskes89PRB, Baskes92PRB, Lee00PRB} is suitable to assess the behavior of \fireo{} with 3-body interactions potentials suitable for complex alloys or covalent material~\cite{Lee07Calphad, Jelinek12PRB,Kim15Calphad,Guenole19SM}.
In \meam, an angular term is added to the energy functional of \eam, making it more suitable for complex materials. Here, \meam is used to model Mg and the complex intermetallics Mg$_{17}$Al$_{12}$ and Mg$_2$Ca.

The Stillinger-Weber (\sw{}) potential \cite{Stillinger85PRB, Vink01JNCS, Pizzagalli13JPCM} is also suitable to assess the behavior of \fireo{} with 3-body interaction potentials, but with a particular focus on covalent materials~\cite{Albenze04PRB,Guenole11AM,Guenole17ASS,Texier19JPCS}. Here, \sw is used for simulating Si.

The \ff by van Beest, Kramer and van Santen (\bks)~\cite{Beest90PRL} is chosen to assess the performance of \fireo{} with long range interactions, in particular electrostatic interactions solved in the reciprocal space~\cite{Leonforte06PRL, Luo15NL}. Here, we use it to model silicate glass, an ionic material that includes long range coulombic interactions.

\subsection{Simulation setups}
\label{sec:simulation-setups}

\begin{figure*}[t]
	\centering
	\includegraphics[width=0.9\textwidth,draft=false]{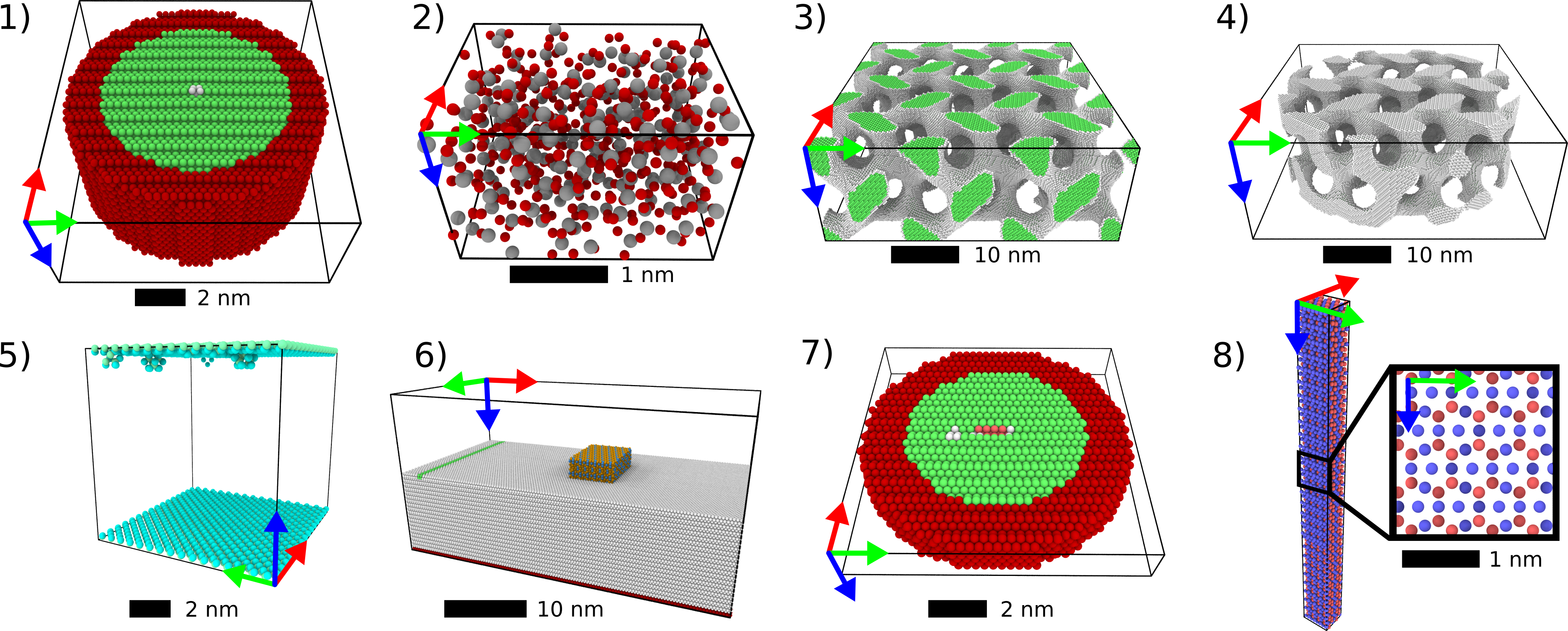}
	\caption{\label{fig:samples}
		Snapshots of the atomistic samples used for the test simulations
		\ref{c:disl}, 
		\ref{c:glass},
		\ref{c:NPGpbc}
		\ref{c:NPGnopbc},
		\ref{c:SiVac},
		\ref{c:Mg},
		\ref{c:AlVac} and
		\ref{c:Laves}.
		Color coding in (\ref{c:disl},\ref{c:NPGpbc},\ref{c:NPGnopbc},	\ref{c:AlVac}) is based on the common neighbor analysis~\cite{Stukowski12MSMSE}: green, \fcc; light red, stacking fault; white, others;
		Color coding in (\ref{c:glass},\ref{c:Laves}) based on chemical species: grey, Si; dark-red, O; blue, Ca; light-red, Mg.
		Color coding in (\ref{c:SiVac}) based on diamond structure analyses~\cite{Maras16CPC}: turquoise, non-diamond atoms; atoms in diamond configuration are removed for clarity.
		Color coding in (\ref{c:Mg}): light-grey, Mg \hcp matrix atoms; green, \fcc dislocation atoms. Within the cuboidale precipitate, orange and blue atoms are Mg and Al, respectively.
		Half of the Mg matrix atoms have been removed for clarity.
		In (\ref{c:disl},\ref{c:AlVac}) dark-red colored atom are frozen.
		The simulation box axis x, y, z are represented by red, green, blue arrows, respectively. The scale bars indicate the dimension of each sample in \textit{nm}.
	}
\end{figure*}

In cases 1--6, the goal is to find a minimum energy configuration starting from some initial state of a system. In cases 7--8 the goal is to find a minimum energy path between two states of the system by \neb. 
The test cases are described in the following and a summary is provided in Tab.~\ref{tab:tests}.
The atomic configurations are illustrated in Fig.~\ref{fig:samples}.

\begin{enumerate}
	\item \label{c:disl} \textbf{Relaxation of a dislocation in Al:} An edge dislocation~\cite{Anderson17} is inserted in an Al cylinder by displacing the atoms according to the anisotropic-elastic solution~\cite{Bacon09DiS}. The cylinder has 25,340 atoms and a radius of 5.2~nm, including a border of width 1.4~nm where atoms are frozen in the $x$ and $y$ directions, see Fig.~\ref{fig:samples}(\ref{c:disl}). Periodic boundary conditions (\pbc) are used in the $z$-direction, with a box length of 5.0~nm. The \eam potential by Mishin~\etal is used~\cite{Mishin99PRB}. 
	
	\item \label{c:glass} \textbf{Relaxation of a 6000K SiO$_2$ melt:} The system consists of 648 atoms (216 Si and 432 O) within a simulation box of $2.0\times2.6\times1.6$~nm$^3$ and \pbc in all directions (Fig.~\ref{fig:samples}(\ref{c:glass})). The melt is obtained by \md from an $\alpha$-quartz crystalline structure.
	Since this configuration is initially far from a $0$~K energy minimum, the maximum atomic displacement per step (\texttt{dmax} in \lmp) had to be set to $0.001$~\AA\ instead of $0.1$~\AA\ (default value).
	This case uses the \bks potential~\cite{Beest90PRL}. The long range coulombic interactions is calculated by a standard Ewald summation with an accuracy of $10^{-5}$ and a direct/reciprocal space cutoff of 1~nm.
	
	\item \label{c:NPGpbc} \textbf{Relaxation of bulk Au with a nano-porous gyroid structure:} The structure has 613,035 atoms and is contained within a box of $44.6\times42.6\times15.7$~nm$^3$ with full \pbc. This case exhibits a particularly high surface over bulk ratio (21.4\% of the atoms belong to surfaces) with complex curvatures, see Fig.~\ref{fig:samples}(\ref{c:NPGpbc}). The \ff is of the EAM type~\cite{Park05PRB}.
	
	\item \label{c:NPGnopbc} \textbf{Relaxation of a Au nano-pillar with a nano-porous gyroid structure:} This case is similar to case \ref{c:NPGpbc}, but without \pbc. The structure consists of 457,424 atoms and has a cylindrical shape with radius 42.6~nm and height 15.7 nm, see Fig.~\ref{fig:samples}(\ref{c:NPGnopbc}). It was cut out of the sample \ref{c:NPGpbc}. 25.6\% of atoms are surface atoms. Due to the absence of periodicity, only surface atoms (white) are visible in Fig.~\ref{fig:samples}(\ref{c:NPGnopbc}).
	
	\item \label{c:SiVac} \textbf{Relaxation of vacancies in Si:} Five vacancies are distributed in a Si slab of 32,762 atoms contained in a box of $8.9^3$~nm$^3$, see  Fig.~\ref{fig:samples}(\ref{c:SiVac}). The $x$ and $y$ directions are periodic. Si is simulated using the \sw \ff with the original parameterization in Ref.~\cite{Stillinger85PRB}.
	
	\item \label{c:Mg} \textbf{Relaxation of a dislocation in Mg with a precipitate:}
	A Mg matrix contains an approximation of the isotropic displacement field of an edge dislocation on one side, and a relaxed 
	Mg\textsubscript{17}Al\textsubscript{12} precipitate on the other side. 
	The Burgers vector of the dislocation is $b = a_0/3 \hkl<-2 1 1 0>$.
	The simulation box of $40\times20\times20$~nm$^3$ contains 694,680 atoms and the precipitate has a cuboidal shape with dimensions of $5.5\times7.8\times6$~nm$^3$, see Fig.~\ref{fig:samples}(\ref{c:Mg}). 
	More details on this setup can be found elsewhere \cite{Vaid19M}.
	The \meam potential from Kim \etal is used~\cite{Kim09Calphad}.
		
	\item \label{c:AlVac} \textbf{Energy barrier for vacancy migration in Al:} \neb is used to calculate the energy barrier for migration of a vacancy near an edge dislocation in Al. The setup is similar to \ref{c:disl}, see Fig.~\ref{fig:samples}(\ref{c:AlVac}). It consists of a cylinder of 7,010 atoms periodic in $z$ direction that contains a vacancy (surrounded by white atoms), and a relaxed  edge dislocation with Burgers vector $1/2a\hkl<110>$ (light-red atoms). The cylinder has a length of 1.5~nm and a radius of 5.0~nm, including a border of width 1.4~nm where atoms are frozen in $x$ and $y$ directions (dark-red atoms). \neb simulations are performed with 6 intermediate configurations, between 2 stable configurations that represent the hopping of the vacancy to a neighboring site. The \ff is the same as in case~\ref{c:disl}.
	
	\item \label{c:Laves} \textbf{Energy barrier of the synchroshear mechanism in Mg\textsubscript{2}Ca:} In brief, the synchroshear mechanism is responsible for the propagation of dislocations in the \hkl{0001} basal plane of the \hcp Laves phase (\textit{Strukturbericht} C14). It involves the synchronous glide of partial dislocations on adjacent basal planes. More details can be found elsewhere~\cite{Vedmedenko08AM, Zhang11PRL,Guenole19SM}.
	The system contains 5,376 atoms in a box of $2.5\times2.1\times28.0$~nm$^3$. \pbc are applied in all directions (See Fig.~\ref{fig:samples}(\ref{c:Laves})). 
	\neb simulations are performed with 18 intermediate configurations as described elsewhere~\cite{Guenole19SM}. 
	The \ff is the \meam from Kim \etal\cite{Kim15Calphad}.
	
\end{enumerate}

\begin{table*}[t]
    \small
	\centering
	\begin{tabular}{p{0.25\textwidth}p{0.25\textwidth}rlrr}
	\hline 
	Case & Specificities & Atoms & \ff & \multicolumn{2}{c}{\fireo{} performance}  \\
	 & & & & \textit{vs} \cgrad & \textit{vs} \fire \\
	\hline
	\ref{c:disl}:
	dislocation in Al &
	\makecell[tl]{Long range  \\ displacement field} &
	25,340  & EAM & 1.2 & 29.3 \\  
	\ref{c:glass}:
	melt of silicate glass &
	\makecell[tl]{
	Electrostatic interactions \\ and local disorder
	}& 
	648 & BKS & $\infty$ & $>3.0 $ \\  
	\ref{c:NPGpbc}:
	nano-porous bulk & 
	Surface tension &
	613,035 & EAM & $(1.5)$ & $\infty$ \\
	\ref{c:NPGnopbc}: 
	nano-porous pillar & 
	\makecell[tl]{Surface tension \\ and free boundaries} &
	457,424 & EAM & $(0.8)$ & $\infty$ \\  
	\ref{c:SiVac}:
	vacancies in silicon &
	\makecell[tl]{3-body force field} &
	32,762 & SW & 1.1 & $>10.0$ \\
	\makecell[tl]{\ref{c:Mg}: 
	dislocation-precipitate \\ interaction} &
	Configuration stability &
	694,680  &  MEAM & $\infty$ & $\infty$ \\ 
	\ref{c:AlVac}:
	vacancy in Al &
	\neb, simple path & 
	7,010 & EAM & -- & 1.8 \\
	\ref{c:Laves}:
	synchroshear  &
	\neb, complex path &
	5,376 & MEAM & -- & 2.9 \\ 
	\hline 
	\end{tabular}
	\caption{Test cases for the implementation of \fireo. The last two columns show the performance of \fireo\ \textit{relative to} \cgrad or \fire, i.e.\ the ratio of forces evaluation required for relaxation: \cgrad/\fireo\ or \fire/\fireo. $\infty$ indicate that \cgrad or \fire\ are much too slow to relax the system, or not able at all. Values in brackets indicate that the relaxation with \cgrad stopped before reaching the threshold (\texttt{line search alpha is zero}) but can still be considered as relaxed.}
	\label{tab:tests}
\end{table*} 

\subsection{Results and discussion}

The computationally most expensive task in atomistic simulations is typically the calculation of the interatomic forces, therefore the number of force evaluations is used for comparing minimizer performances.
Except otherwise mentioned, the threshold \fnorm used in this work is $=10^{-8}$eV/\AA.
The evolution of \texttt{f2norm} as a function of the number of force evaluations is shown in Fig.~\ref{fig:samples}. Tab.~\ref{tab:tests} indicates the increase in performance obtained by \fireo \textit{versus} \cgrad and \fire. The performance in optimizing a configuration is determined by the ratio of the number of forces evaluations required by \cgrad or \fire to reach the threshold, over the number of forces evaluations required by \fireo. 
A comparison with \lbfgs is outside the scope of this work which is based on \lmp, where \lbfgs is not included.  A recent comparison between a \fire-based algorithm and \lbfgs was reported by Shuang \etal\cite{Shuang19CMS}.

\begin{figure*}
	\centering
	\begin{subfigure}[t]{0.45\textwidth}
		\caption{Dislocation in aluminum.
		}
		\includegraphics[width=\linewidth]{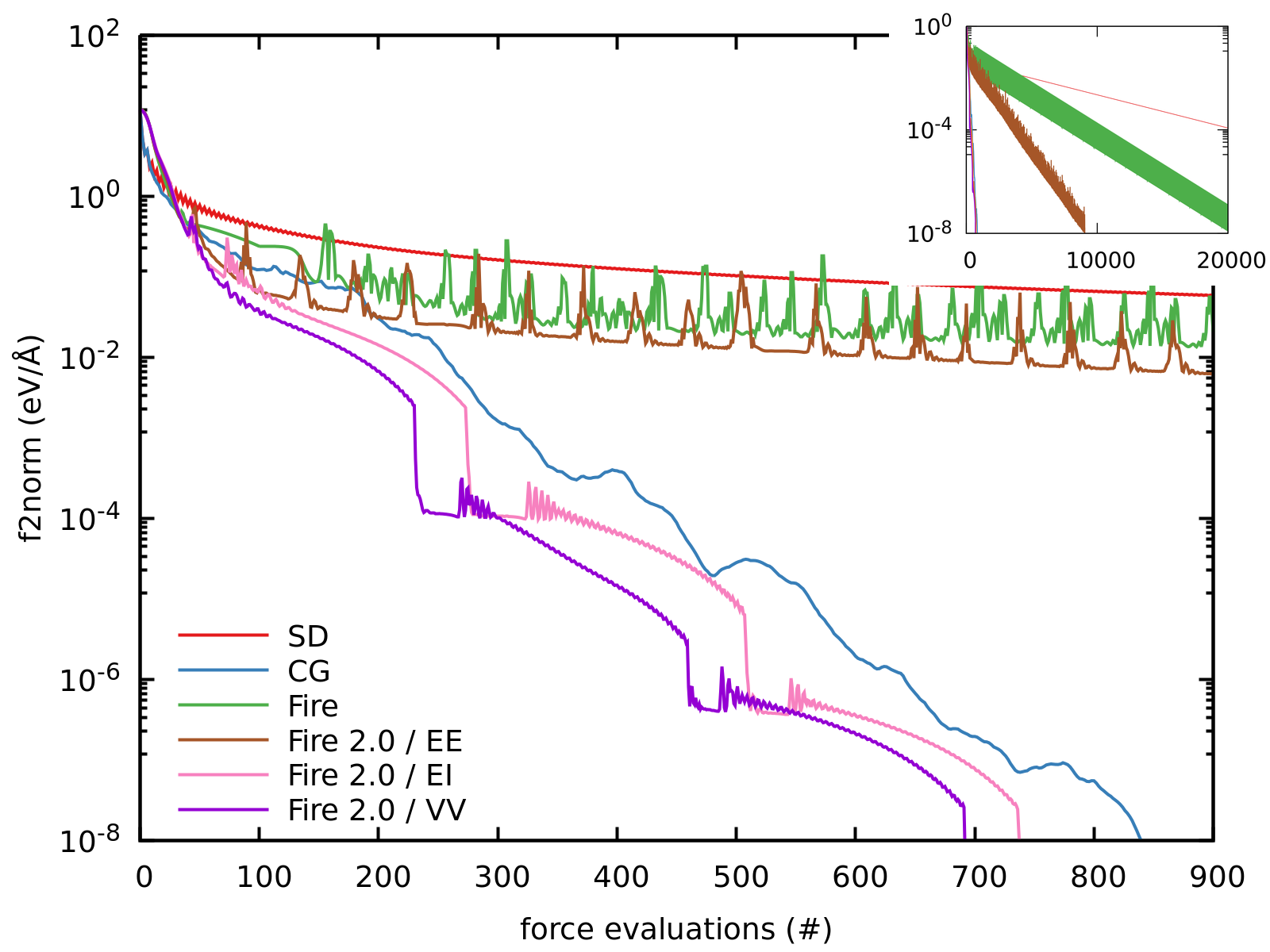} 
		\label{fig:disloc}
	\end{subfigure} 
	\begin{subfigure}[t]{0.45\textwidth}
		\caption{Silica melt.
		}
		\includegraphics[width=\linewidth]{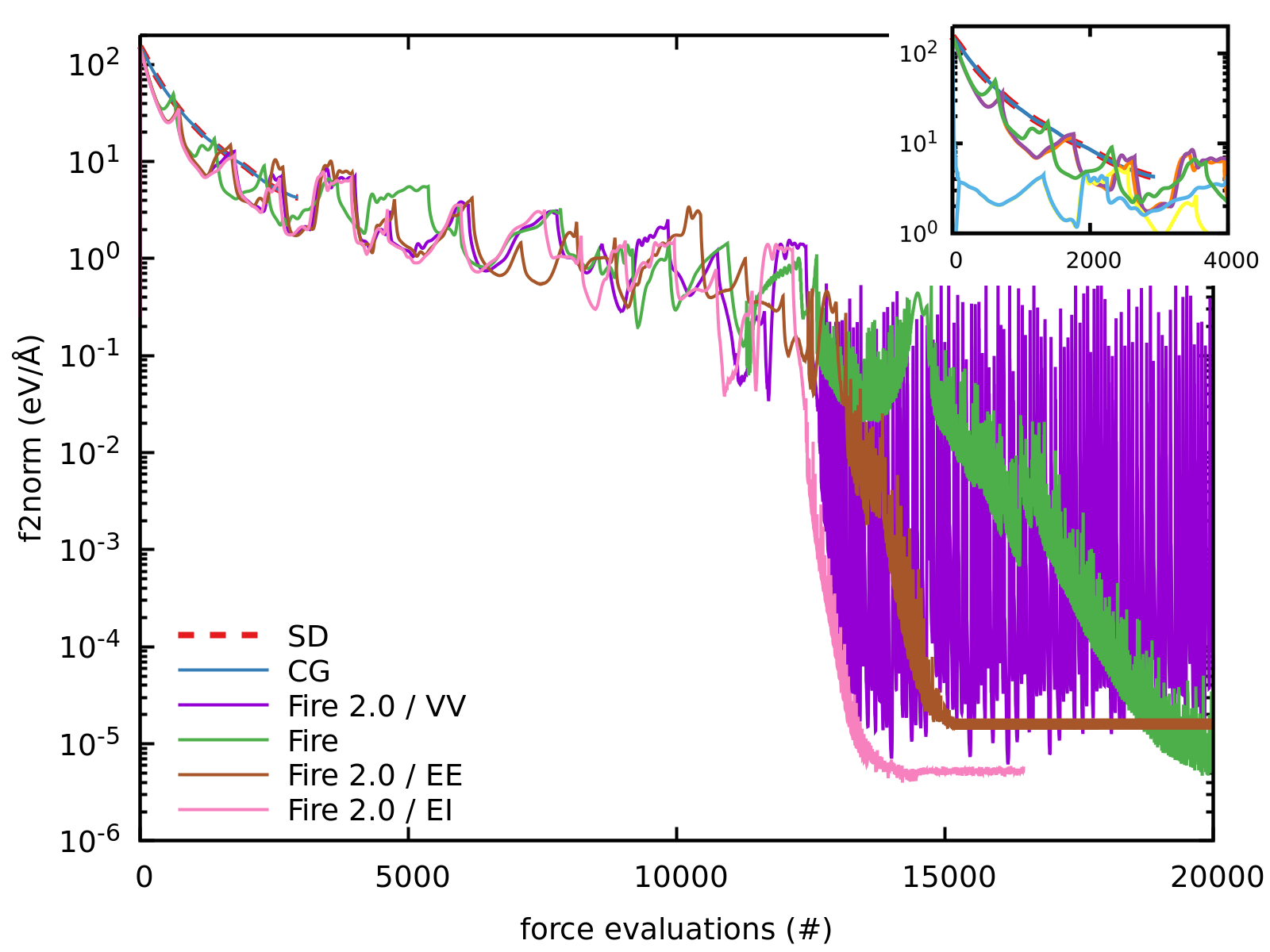} 
		\label{fig:glassbks} 
	\end{subfigure}

	\begin{subfigure}[t]{0.45\textwidth}
		\caption{Nano-porous gold, bulk.
		}
		\includegraphics[width=\linewidth]{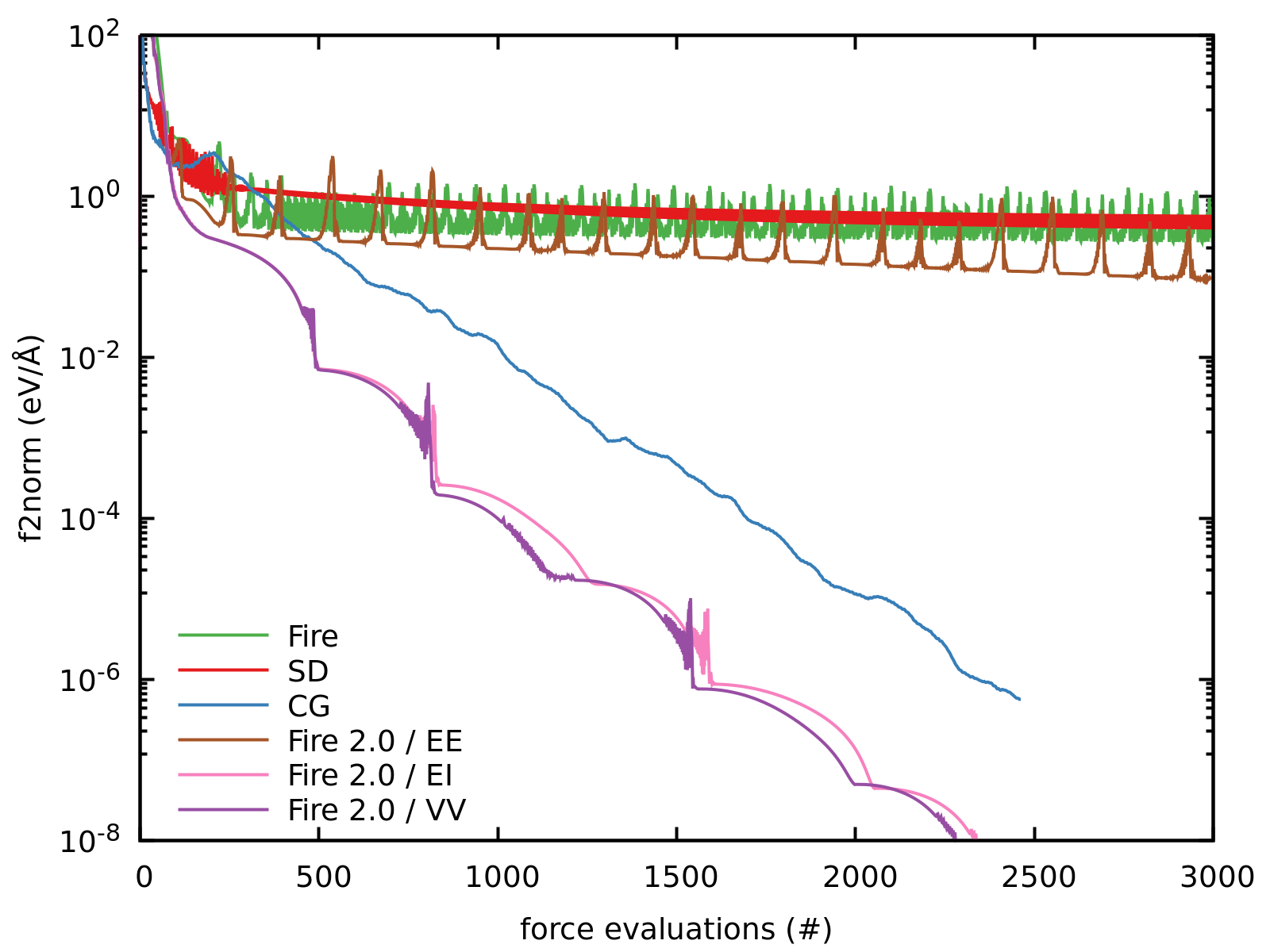} 
		\label{fig:NPGpbc} 
	\end{subfigure}
	\begin{subfigure}[t]{0.45\textwidth}
		\caption{Nano-porous gold, pillar.
		}
		\includegraphics[width=\linewidth]{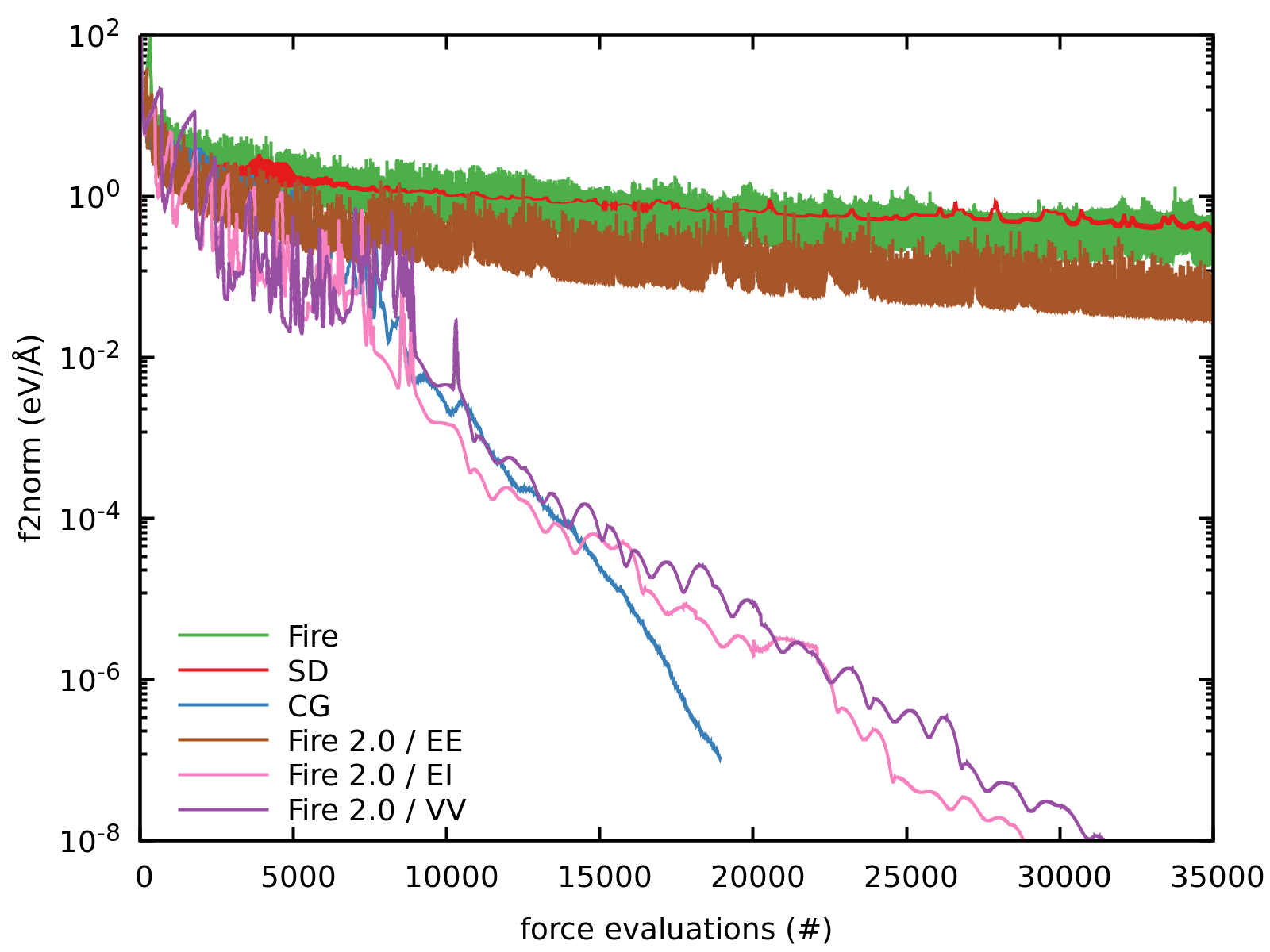} 
		\label{fig:NPGnopbc} 
	\end{subfigure}
	
	\begin{subfigure}[t]{0.45\textwidth}
		\caption{Silicon vacancies.
		}
		\includegraphics[width=\linewidth]{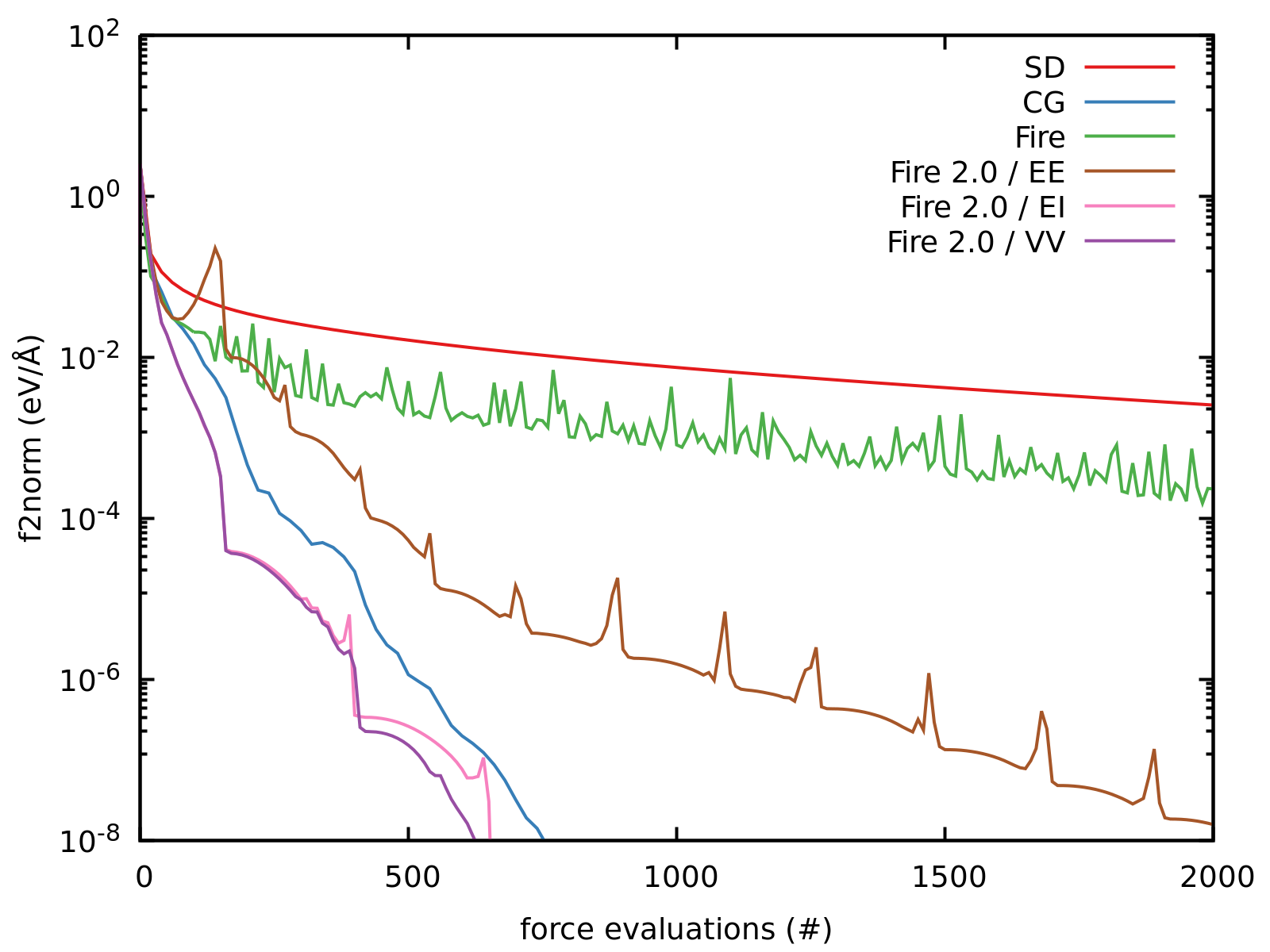} 
		\label{fig:sivacancies} 
	\end{subfigure}
	\begin{subfigure}[t]{0.45\textwidth}
		\caption{Dislocation and precipitate in Mg alloys.
		}
		\includegraphics[width=\linewidth]{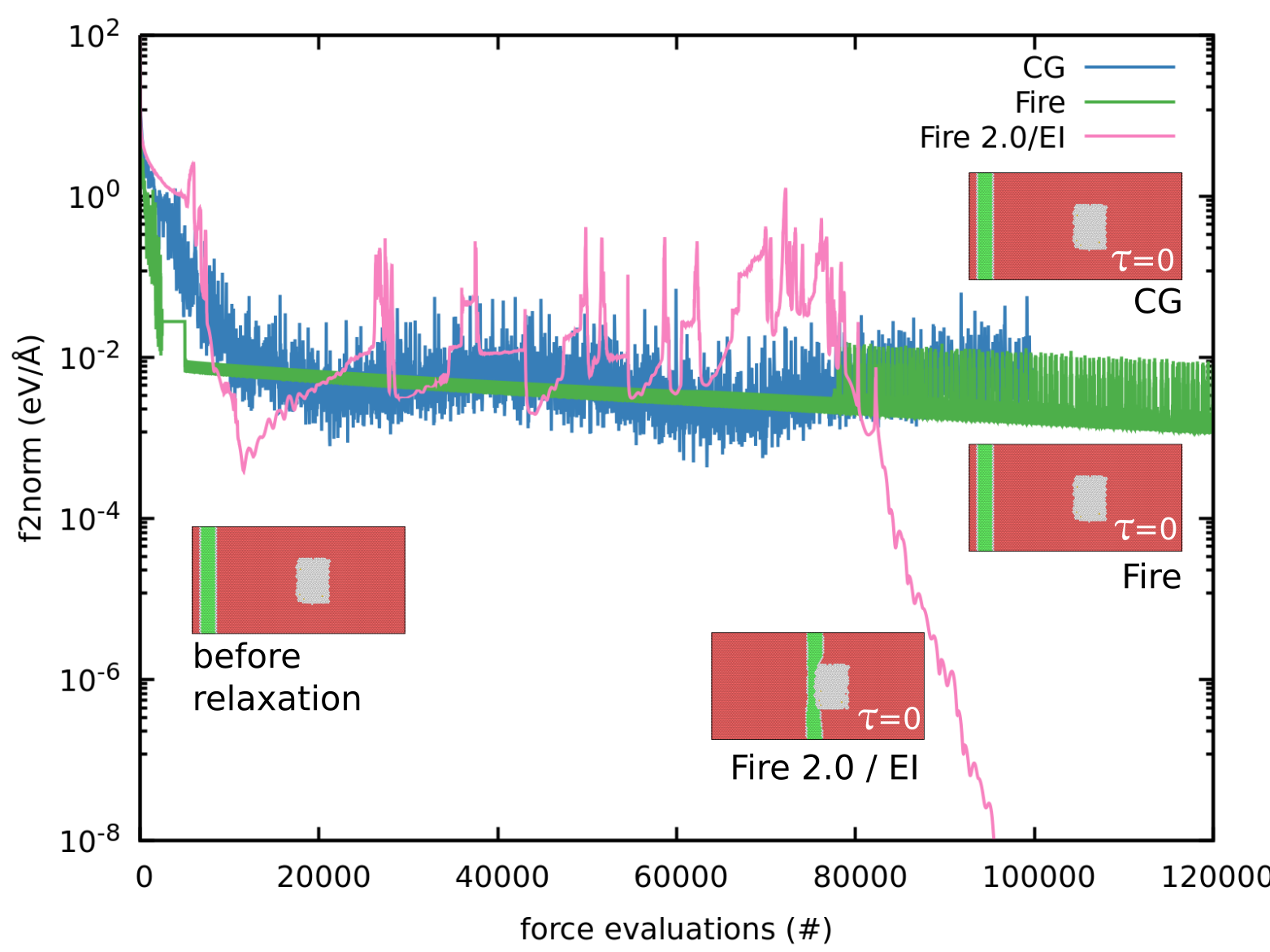} 
		\label{fig:mgalloy} 
	\end{subfigure}
	\caption{Force \texttt{f2norm} as a function of the number of interatomic forces evaluation during minimization. Subfigures 1 to 6 correspond to the test cases 1 to 6, respectively.
	(Continue on next page.)
	}
	\label{fig:f2norm}
\end{figure*}
\addtocounter{figure}{-1}
\begin{figure*} [t!]
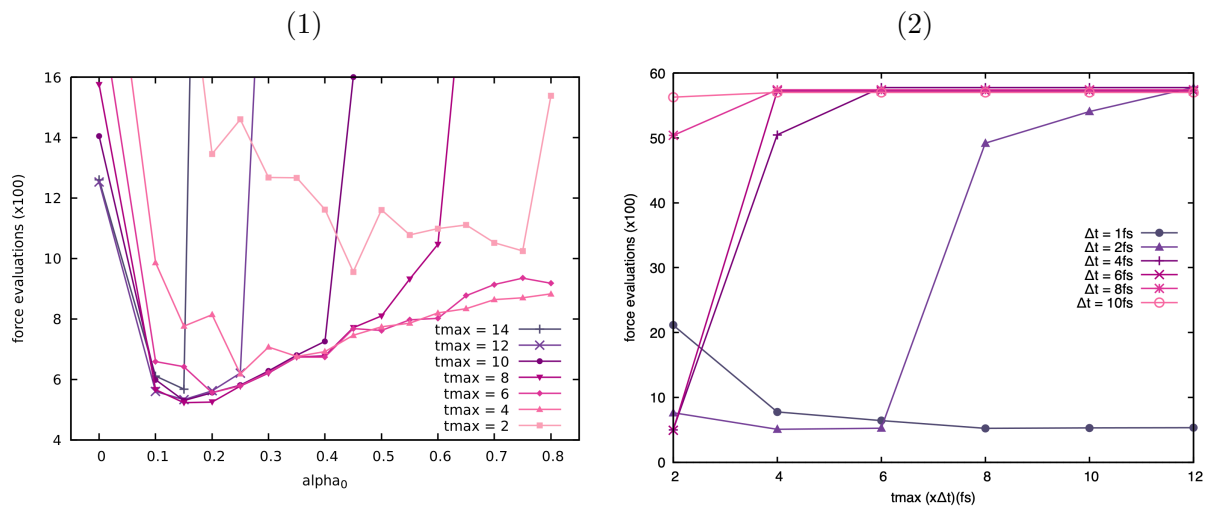

	\caption{(Continued) The color of curves indicates the minimization method: steepest descent (\sdesc, red line), conjugate gradient (\cgrad, blue line), \fire (Fire, green line). For \fireo, the color indicates the time integration scheme: Euler Explicit (EE, brown line), Euler Semi-implicit (EI, pink line) and Velocity-Verlet (VV, purple line).
	Insets in \subref{fig:mgalloy} represent the minimized configurations for different minimization methods, with atoms colored according to the common neighbors analysis method (red, Mg HCP; green, Mg FCC and dislocation; grey, Mg\textsubscript{17}Al\textsubscript{12} precipitate).
	}
\end{figure*}

\subsubsection{\cgrad vs \fireo}

\fireo\ performs better than \cgrad in the two simple cases \ref{c:SiVac} and \ref{c:disl}, with a ratio of $1.1\times$ and $1.2\times$, respectively.
The relaxation of the long range \ff in the case \ref{c:glass} is not possible using \cgrad, which terminates with the \lmp's stopping criterion \texttt{linesearch alpha is zero} at comparatively large \texttt{f2norm}. Generally, this occurs when no minimum can be found by line search, for example when the backtracking algorithm backtracks all the way to the initial point.
A similar behavior is seen in test case \ref{c:Mg}: \cgrad fails to reduce the forces sufficiently.
Note that the output configuration is clearly different to the one obtained with \fireo{}, see the insets in Fig.~\ref{fig:mgalloy}. Similarly to \fire, \cgrad predicts that the dislocation remains in the Mg matrix far away from the precipitate, whereas \fireo\ predicts that the dislocation moves towards the precipitate.

In test cases \ref{c:NPGpbc} and \ref{c:NPGnopbc} (nano-porous Au) \cgrad{} fails to reach the strict \fnorm threshold of $10^{-8}$eV/\AA, see Figs.\ \ref{fig:NPGpbc} and \ref{fig:NPGnopbc}. Line search fails when \fnorm is below $10^{-6}$eV/\AA. To quantify the performance of \fireo{}, we thus compare the number of force evaluations requires to reach the lowest \fnorm achieved by \cgrad.  Here, \fireo{} performs better than \cgrad{} in problem \ref{c:NPGpbc} (bulk), but worse in \ref{c:NPGnopbc} (free boundaries).

\subsubsection{\fire vs \fireo}

\fireo performs better than \fire as implemented in \lmp in all the test cases. The smallest speedups of $1.8\times$ and $2.9\times$ are seen in the \neb calculations of problems \ref{c:AlVac} and \ref{c:Laves}, respectively. A larger speedup of more than $3\times$ is obtained in case \ref{c:glass}, where the \bks potential is used. Note that it is particularly difficult to relax the long-range coulombic interaction, so that the desired force stopping criterion was not reached. 
The relaxation with \fireo\ stopped when \texttt{f2norm} reached a plateau, see Fig.~\ref{fig:glassbks}. In this plateau region, \fireo{} detected repeated attempts at uphill motion ($P(t)<0$), and so minimization was terminated with return value \texttt{MAXVDOTF}. 
A speedup can still be defined by comparing the number of force evaluations at which  \fire\ reaches a \texttt{f2norm} similar to the one at the end of the \fireo{} minimization.  Much larger speedups of $10\times$ and $30\times$ are obtained in the cases \ref{c:SiVac} and \ref{c:disl}, respectively. Finally, in the cases \ref{c:NPGpbc},  \ref{c:NPGnopbc} and \ref{c:Mg} convergence of \fire{} is so slow that the
desired \texttt{f2norm} threshold is not reached.

\subsubsection{\fireo: Influence of the time integration scheme}

Fig.~\ref{fig:f2norm} shows the minimization of the problems \ref{c:disl} to \ref{c:Mg} with \fireo and the four integration schemes. 
With an Euler Explicit scheme, \fireo{} shows a similar poor performance as \sdesc{} and \fire. 
Switching to Euler Implicit integration improves the performance significantly. 
With this integrator, \fireo\ typically outperforms \cgrad in all these cases. 
The Velocity Verlet integrator, on the other hand, performs slightly better than the others in problems \ref{c:disl}, \ref{c:NPGpbc} and \ref{c:SiVac}, but not in problems \ref{c:glass} and \ref{c:NPGnopbc}.
In particular, the case \ref{c:glass} (Fig.~\ref{fig:glassbks}) has stability issues.

As for the cases \ref{c:disl} to \ref{c:Mg}, the \neb cases \ref{c:AlVac} and \ref{c:Laves} show a similar poor performance as \fire while using \fireo with an Euler Explicit scheme. By switching to Euler Implicit integration, as before, \fireo typically outperforms \fire. The Velocity Verlet integrator exhibits mixed behavior: it performs better than the other integrators in problem \ref{c:AlVac} but not in problem \ref{c:Laves}, the latter having stability issues.

\subsubsection{\fireo: Influence of individual parameters}

We have investigated the influence of the parameters $\alpha_{start}$ and $\Delta t_{max}$ on the performance of \fireo. Since the observed trends do not depend on the problem, the computationally less expensive problem~\ref{c:SiVac} has been chosen for this parameter study.
Note that $\alpha_{start}$ and $\Delta t_{max}$ are controlled by the \lmp parameters \texttt{alpha0} and \texttt{tmax}, respectively. 

\begin{figure*}
	\begin{subfigure}[t]{0.48\textwidth}
		\caption{
		}
		\includegraphics[width=\linewidth]{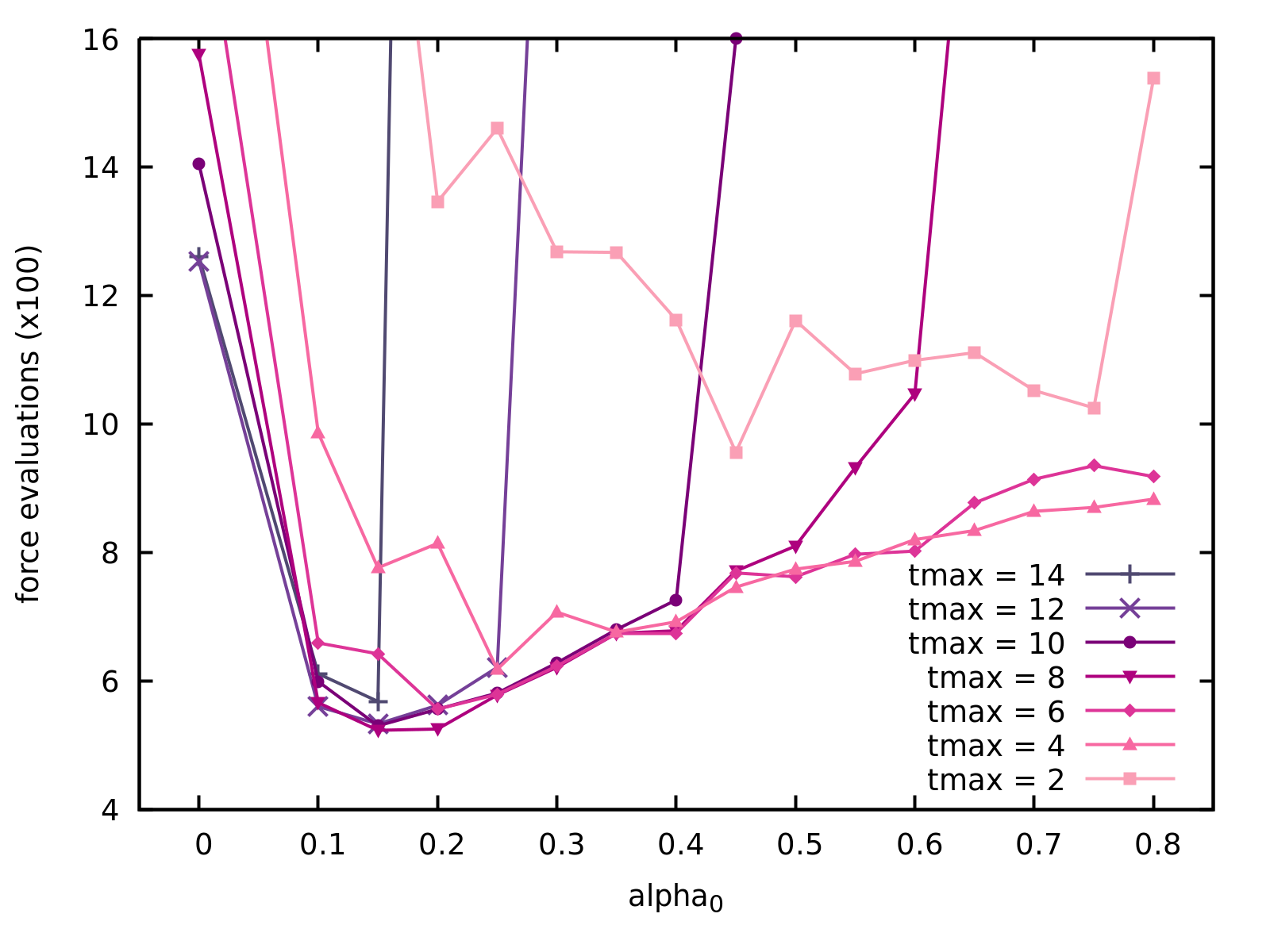} 
		\label{fig:alpha0}
	\end{subfigure} 
	\begin{subfigure}[t]{0.48\textwidth}
		\caption{
		}
		\includegraphics[width=\linewidth]{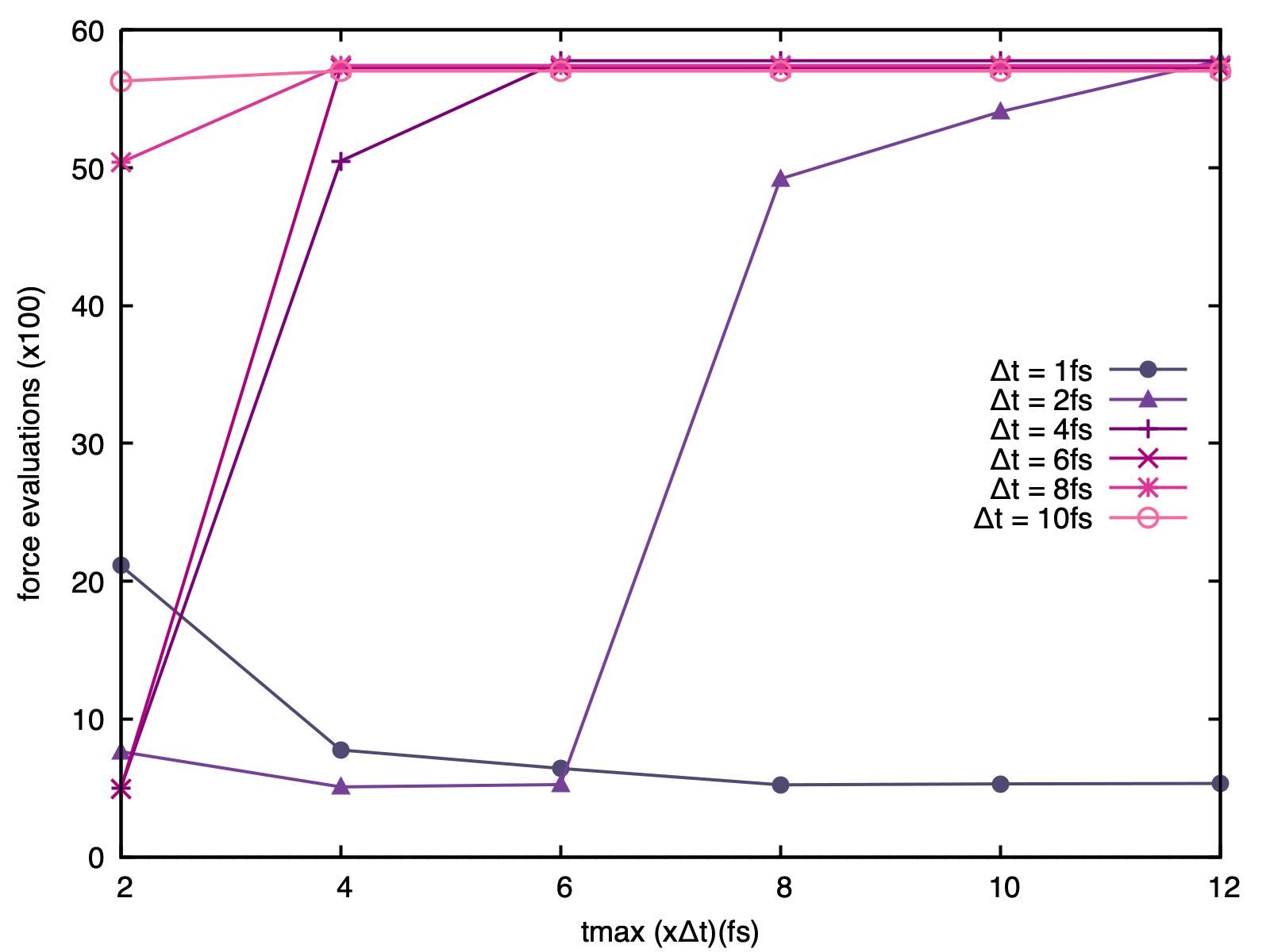} 
		\label{fig:tmax} 
	\end{subfigure}
	\caption{Influence of the parameters \texttt{alpha0} (1) and \texttt{tmax} (2) on the minimization performances, characterized by the number of force evaluations required to reach the force threshold in the case~\ref{c:SiVac}.
	(1) shows the performance as a function of \texttt{alpha0} for different choice of \texttt{tmax}, with $\Delta$t~=~1fs. (2) shows the performance as a function of \texttt{tmax} for different choice of \texttt{timestep} $\Delta$t, with \texttt{alpha0} = 0.15. 
	}
	\label{fig:param}
\end{figure*}

The performance as a function of $\alpha_{start}$ for different choice of $\Delta t_{max}$ ($\Delta t = 1$ps) are shown on Fig.~\ref{fig:alpha0}.
As seen on Fig.~\ref{fig:alpha0}, best values lie in a range from 0.10 to 0.25. Increasing $\alpha_{start}$ does not improve performance. For $\Delta t_{max} > 6$ps, it even leads to dramatic performance reduction. 
Generally, large values of \texttt{tmax} will benefit from lower values of \texttt{alpha0}, around 0.10 -- 0.15.

The performances as a function of $\Delta t_{max}$ for different choice of $\Delta t$ ($\alpha_{start} = 0.15$) is shown on Fig.~\ref{fig:tmax}.
The maximum value for the timestep $\Delta t_{max}$ is controlled by the coefficient \texttt{tmax} applied on the timestep $\Delta t$. That is $\Delta t_{max} = \texttt{tmax} * \texttt{timestep}$.
As seen on Fig.~\ref{fig:tmax}, the performances largely depend on the correlated choice of the timestep and $\texttt{tmax}$. The optimum $\Delta t$ for running dynamics simulation in such system being \SI{1}{\femto\second}, it appears that choosing $\Delta t$ at least 4 times bigger is not relevant and leads to poor performances. In this case, \fireo\ shows good performance for $\Delta t_{max} <  \SI{12}{\femto\second}$, which correspond to \texttt{tmax} from 6 to 12, depending on the timestep. 
Generally, one can consider reducing \texttt{tmax} to improve the stability of the minimization.

\subsubsection{\fireo: Nudged elastic band method}

\fireo{} is $1.8$ and $2.9$ times faster than \fire in the cases  \ref{c:AlVac} and \ref{c:Laves}, respectively, see Tab.~\ref{tab:tests}. Note that case \ref{c:Laves}, where the relative performance of \fireo{} is better,  is also the more complex case (complex mechanism and more images). Finally, a
comparison of \fireo and \cgrad{} is not possible in these cases, because \neb calculations in \lmp{} require damped dynamics minimizers.

\subsubsection{\fireo: On the usage of preconditioners}

For geometrical optimization of atomistic configurations, preconditioners are known to largely enhanced the efficiency of the algorithms by considering  known characteristics of the system, like the local atomic neighborhood~\cite{Schlick10Book}. For more details on preconditioners and how to determine them, the reader should refer to the recent work of Packwood \etal\cite{Packwood16JCP}. 
Preconditioners are especially efficient on large systems and could then reduce the difference we observe between \cgrad and \fireo.
With a similar goal as the preconditioner, that is the reduction of degrees of freedom to be optimized, we also investigated the influence of a pre-relaxation with a different minimizer on the performance of \fireo. This pre-relaxation was performed with the \texttt{quickmin} minimizer for 100 iterations, as implemented in \lmp~\cite{Sheppard08JCP}. 
In all the problems but one, we did not observe any gain. Only the case \ref{c:glass} with long range atomic interactions evidence a significant advantage of performing this pre-relaxation, with a speedup close to 30\%.
That also improved the stability of \fireo with Velocity-Verlet for the same problem. 

\subsection{General aspects}

\fireo minimizes faster than \fire and can potentially reach lower residual forces.
In the case of \neb simulations, the performance gain increases with the complexity of the setup.
When comparing to \cgrad, \fireo\ shows better performance except in case \ref{c:NPGnopbc}, the non-periodic nanoporous Au structure (Fig.~\ref{fig:NPGnopbc}). Recall that the system was created by cutting bulk nanoporous Au (case~\ref{c:NPGpbc}) and removing \pbc. The structure thus undergoes a sudden global shrinkage at the beginning of the minimization, which can easily be optimized by \cgrad. 
In contrast, pseudo-dynamics relaxators like \fire\ and \fireo are sensitive to such scaling by generating a shock wave that has to be damped during optimization and thus may hamper minimization.
In the bulk case (case~\ref{c:NPGpbc}), where there is no such global dynamic effect, \fireo{} performs better than \cgrad{}, which also indicates that the algorithm remains robust with a large amount of free surfaces.
On all other systems, \fireo\ shows various levels performance increase in comparison to \cgrad, between 20\% and 3000\%.
In addition, \cgrad is not able to minimize the forces in some cases, due either to the long range stress field (case~\ref{c:Mg}) or long range atomic interactions (case~\ref{c:glass}).

\cgrad{} sometimes terminates prematurely (at a high level of residual force), because line search fails. Similarly, \fire{} or \fireo{} could terminate prematurely if convergence is slow and the chosen maximum number of force evaluations is too low.  The resulting structure is then insufficiently optimized. Here, this was seen in case~\ref{c:Mg} (Fig.~\ref{fig:mgalloy}), where \fire and \cgrad yield a different dislocation position than \fireo. The latter is less susceptible to premature termination, because it does not suffer from line search problems and typically minimizes with fewer number of force evaluations. 
As a general statement, we note that reaching low \texttt{f2norm} values is crucial and analyzing an insufficiently relaxed structure could lead to wrong interpretations. This is especially important in statics and quasi-statics calculations of critical stresses. As a good practice, we suggest to always indicate the exact \texttt{f2norm} value alongside results in published work. 

Among all the parameters that affect the behavior and performance of \fireo, the time integration scheme is the most important. 
Overall, as presented in the results above, Euler Implicit integrator provides robust minimizations at the cost of a slightly reduce performance. Hence we recommend the usage of \fireo\ with an Euler Implicit integrator. 
Similarly, the very recent work of Shuang \etal\cite{Shuang19CMS} also recommended to couple the \fire approach with a semi-implicit Euler integrator.

More generally, Tab.~\ref{tab:parameters} list the parametrization of \fireo\ accessible by the command \texttt{min\_modify} as implemented in \lmp  and the associated default values we recommend to use.  
More specifically, \texttt{tmax} can be reduced to improve the stability but should range from $2$ to $12$, and \texttt{alpha0} should range from $0.10$ to $0.25$. In any case, we recommend to set the simulation timestep (command \texttt{timestep} in \lmp) to the same value as in MD at low temperature.

\section{Summary}

In this work we describe \fireo, an optimized version of the \fire minimization algorithm within the \lmp molecular dynamics simulator, and add important details to the canonical publication~\cite{Bitzek06PRL}.
The choice of time integration scheme has appeared to be crucial for \fire and is now clearly discussed. A non-symplectic scheme like \textit{Euler explicit} should not be used.
We have shown the clear advantages of \fireo \textit{versus} \fire and \textit{versus} conjugate gradient through several examples in materials science:
\fireo is significantly faster than \fire or conjugate gradient and can result in lower energy structures not found by other algorithms.

We intend \fireo to entirely replace \fire, the present work being a complement of the original publication~\cite{Bitzek06PRL}.
Ultimately, this work intends to provide insights on performing more accurate and more efficient forces minimization of atomistic systems.

\section*{Acknowledgments}
J.G is thankful for the financial support by the German Research Foundation (DFG) through the priority program SPP 1594 “Topological Engineering of Ultra-Strong Glasses”.
F.H acknowledges financial support by the DFG through projects C3 (atomistic simulations) of SFB/Transregio 103 (Single Crystal Superalloys).
Z.X. acknowledges financial support by the German Science Foundation (DFG) via the research training group GRK 1896 “In Situ Microscopy with Electrons, X-rays and Scanning Probes”.
W.N. acknowledges financial supports by the European Union, within the starting grant ShapingRoughness (757343) and the advanced grant PreCoMet (339081). 
EB gratefully acknowledges the funding from European Research Council (ERC) through the project “microKIc” (grant agreement No. 725483).
A.V., A.P. and E.B. acknowledges the support of the Cluster of Excellence Engineering of Advanced Materials (EAM).
A.P. and E.B. acknowledges the support of the Central Institute of Scientific Computing (ZISC).
Computing resources were provided by the Regionales RechenZentrum Erlangen (RRZE) and by RWTH Aachen University under project rwth0297 and rwth0407.
The authors gratefully thank Jim Lutsko, University Libre de Bruxelles, for helpful discussions on this manuscript.

\section*{Data availability}
The source code of the implementation of \fireo in \lmp is freely available online, as described in \ref{sec:source}.
The raw data required to reproduce the findings presented in this paper cannot be shared at this time as the data also forms part of an ongoing study.

\bibliographystyle{elsarticle-num}

\newpage
\appendix
\section{Integration in \fireo{}}

\begin{algorithm}
\caption{Explicit Euler integration in \fireo{}}\label{algo:eulerexpl}
\begin{algorithmic}[1]              
\State $\mathbf{v}(t) \gets 
(1 - \alpha) \cdot \mathbf{v}(t) 
+ \alpha  \mathbf{F}(\mathbf{x}(t)) \cdot |\mathbf{v}(t)|/\vert\mathbf{F}(\mathbf{x}(t))\vert$
\Comment{Mixing}
\State $\mathbf{x}(t+\Delta t) \gets \mathbf{x}(t) + \Delta t \cdot \mathbf{v}(t) $
\State $\mathbf{v}(t+\Delta t) \gets \mathbf{v}(t) + \Delta t \cdot \mathbf{F}(\mathbf{x}(t+\Delta t)) / m$
\State Calculate $E(x(t+\Delta t))$ 
\State $\mathbf{F}(\mathbf{x}(t+\Delta t)) \gets -\vec{\nabla}E(\mathbf{x}(t+\Delta t))$
\end{algorithmic}
\end{algorithm}

\begin{algorithm}
\caption{Semi-implicit Euler integration in \fireo{}}\label{algo:eulerimpl}
\begin{algorithmic}[1]              
\State $\mathbf{v}(t+\Delta t) \gets \mathbf{v}(t) + \Delta t \cdot \mathbf{F}(\mathbf{x}(t)) / m$
\State $\mathbf{v}(t+\Delta t) \gets 
 (1 - \alpha) \cdot \mathbf{v}(t) 
    + \alpha  \mathbf{F}(\mathbf{x}(t)) \cdot |\mathbf{v}(t)|/\vert\mathbf{F}(\mathbf{x}(t))\vert$
    \Comment{Mixing}
\State $\mathbf{x}(t+\Delta t) \gets \mathbf{x}(t) + \Delta t \cdot \mathbf{v}(t+\Delta t) $
\State Calculate $E(x(t+\Delta t))$ 
\State $\mathbf{F}(\mathbf{x}(t+\Delta t)) \gets -\vec{\nabla}E(\mathbf{x}(t+\Delta t))$
\end{algorithmic}
\end{algorithm}

\begin{algorithm}
\caption{Leapfrog integration in \fireo{}}\label{algo:leapfrog}
\begin{algorithmic}[1]
\State $\mathbf{v}(-\nicefrac{1}{2}\Delta t) \gets 
- \nicefrac{1}{2}\Delta t \cdot \mathbf{F}(\mathbf{x}(t)) / m$
\Comment{Initialization}
\State $\mathbf{v}(t+\nicefrac{1}{2}\Delta t) \gets \mathbf{v}(t-\nicefrac{1}{2}\Delta t) + \Delta t \cdot \mathbf{F}(\mathbf{x}(t)) / m$
\State $\mathbf{v}(t+\nicefrac{1}{2}\Delta t) \gets 
(1 - \alpha) \cdot \mathbf{v}(t+\nicefrac{1}{2}\Delta t) 
+ \alpha  \mathbf{F}(\mathbf{x}(t)) \cdot 
|\mathbf{v}(t+\nicefrac{1}{2}\Delta t)|/\vert\mathbf{F}(\mathbf{x}(t))\vert$
\Comment{Mixing}
\State $\mathbf{x}(t+\Delta t) \gets \mathbf{x}(t) + \Delta t \cdot \mathbf{v}(t+\nicefrac{1}{2}\Delta t) $
\State Calculate $E(x(t+\Delta t))$ 
\State $\mathbf{F}(\mathbf{x}(t+\Delta t)) \gets -\vec{\nabla}E(\mathbf{x}(t+\Delta t))$
\end{algorithmic}
\end{algorithm}

\begin{algorithm}
\caption{Velocity Verlet integration in \fireo{}}\label{algo:verlet}
\begin{algorithmic}[1]              
\State $\mathbf{v}(t+\nicefrac{1}{2}\Delta t) \gets 
\mathbf{v}(t) + \nicefrac{1}{2}\Delta t \cdot \mathbf{F}(\mathbf{x}(t)) / m$
\State $\mathbf{v}(t+\nicefrac{1}{2}\Delta t) \gets 
(1 - \alpha) \cdot \mathbf{v}(t+\nicefrac{1}{2}\Delta t) 
+ \alpha  \mathbf{F}(\mathbf{x}(t)) \cdot 
|\mathbf{v}(t+\nicefrac{1}{2}\Delta t)|/\vert\mathbf{F}(\mathbf{x}(t))\vert$
\Comment{Mixing}
\State $\mathbf{x}(t+\Delta t) \gets \mathbf{x}(t) + \Delta t \cdot \mathbf{v}(t+\nicefrac{1}{2}\Delta t) $
\State Calculate $E(x(t+\Delta t))$ 
\State $\mathbf{F}(\mathbf{x}(t+\Delta t)) \gets -\vec{\nabla}E(\mathbf{x}(t+\Delta t))$
\State $\mathbf{v}(t+\Delta t) \gets 
\mathbf{v}(t+\nicefrac{1}{2}\Delta t) + \nicefrac{1}{2}\Delta t \cdot \mathbf{F}(\mathbf{x}(t+\Delta t)) / m$
\end{algorithmic}
\end{algorithm}

\newpage
\section{Source code of \fireo{}}
\label{sec:source}

\fireo is currently being pulled in the \texttt{master} branch of \lmp (\url{https://github.com/lammps/lammps/pull/1052}) and should replace the current implementation of \fire. Please refer to the documentation of \lmp for the most up-to-date indications.

The development version of the source code is available as a fork of \lmp.  It can be found in the GitHub repository of JG, branch \texttt{adaptglok} (\url{https://github.com/jguenole/lammps/tree/adaptglok}).
The latest commit to the date of this manuscript is \texttt{426ca97} (\url{https://github.com/lammps/lammps/pull/1052/commits/426ca97aa6ecf289f824a70781f3640d429e6ab3}).

\end{document}